\documentclass[lettersize,journal]{IEEEtran}
\usepackage{amsmath,amsfonts}
\usepackage{algorithmic}
\usepackage{algorithm}
\usepackage{array}
\usepackage[caption=false,font=normalsize,labelfont=sf,textfont=sf]{subfig}
\usepackage{textcomp}
\usepackage{stfloats}
\usepackage{url}
\usepackage{verbatim}
\usepackage{graphicx}
\usepackage{cite}
\begin{document}

\title{Extremum Seeking Control Based Adaptive Compensation of Position Sensor Harmonics in PMSM Drives}

\author{%
Gayan V. Dissanayake,~\IEEEmembership{Student Member,~IEEE} and
Sandun S. Kuruppu,~\IEEEmembership{Senior Member,~IEEE}%
\thanks{Gayan V. Dissanayake and Sandun Kuruppu are with Western Michigan
University. 4601 Campus Drive, Kalamazoo, MI 49008, USA. Email:
\\g.v.dissanayakemudiyanselage@wmich.edu, sanduns@ieee.org}
}



\maketitle

\begin{abstract}

Permanent Magnet Synchronous Machines (PMSMs) have become one of the preferred forms of electromechanical energy converters, attributing to their high efficiency, torque density, and other unique advantages. However, given the need for proper rotor position measurement for commutation and field orientation, accurate rotor position sensing is of paramount importance. In sensing motor rotor position with a sensor, harmonic errors that arise in the sensing subsystem lead to undesirable torque ripple. Thus, this paper presents an adaptive, extremum-seeking control-based approach capable of mitigating position signal harmonics in PMSMs. The proposed approach is experimentally validated under varying torque, speed, and harmonic conditions. Its harmonic compensation performance is comparatively evaluated against the look-up table–based method. Furthermore, the accuracy of the proposed approach is analyzed, highlighting its effectiveness.

\end{abstract}


\section*{Nomenclature}

\begin{IEEEdescription}

\item[$\theta_m$] Measured rotor position
\item[$\theta_r$] True rotor position
\item[$\theta_c$] Corrected rotor position
\item[$\Delta \theta$] Position sensor offset error
\item[$A_k$] Amplitude of the $k^{\mathrm{th}}$ harmonic
\item[$\phi_k$] Phase angle of the $k^{\mathrm{th}}$ harmonic
\item[$T_{em}$] Electromagnetic torque
\item[$P$] Number of poles of the motor
\item[$\lambda_{m}^{\prime r}$] Flux linkage constant of the PMSM
\item[$I_{qs}^{r}$] Measured q-axis current in the rotor reference frame
\item[$I_{qs}^{r\ast}$] Reference q-axis current in the rotor reference frame
\item[$I_{ds}^{r}$] Measured d-axis current in the rotor reference frame
\item[$I_{ds}^{r\ast}$] Reference d-axis current in the rotor reference frame
\item[$L_d$] Stator d-axis inductance of the PMSM
\item[$L_q$] Stator q-axis inductance of the PMSM
\item[$r_s$] Stator phase resistance of the PMSM
\item[$\omega_r$] Electrical angular frequency of the motor
\item[$V_{qs}^{r}$] Actual q-axis voltage in the rotor reference frame
\item[$V_{ds}^{r}$] Actual d-axis voltage in the rotor reference frame
\item[$y$] Output to be minimized
\item[$f^{\ast}$] Minimum or maximum value of the function
\item[$f^{\prime\prime}$] Second derivative of the function $f(\theta)$
\item[$\theta^{\ast}$] Extremum point corresponding to the minimum or maximum of the function
\item[$\hat{\theta}$] Estimated value of $\theta^{\ast}$
\item[$B$] Integrator gain of the ESC algorithm
\item[$A$] Amplitude of the perturbation signal
\item[$\omega_{esc}$] Perturbation frequency of the ESC algorithm
\item[$\omega_{c1}$] Cutoff frequency of high-pass filter 1
\item[$\omega_{c2}$] Cutoff frequency of high-pass filter 2
\item[$I_a$] Phase-A current of the PMSM
\item[$I_b$] Phase-B current of the PMSM
\item[$I_c$] Phase-C current of the PMSM

\end{IEEEdescription}

\section{Introduction}
\IEEEPARstart{P}{ermanent} Magnet Synchronous Machines (PMSMs) are widely used across various applications due to their exceptional combination of high efficiency, compact size, and superior torque density. These features stem from the permanent magnets, which enable high performance and improved thermal management compared to other motor types. Among these applications, safety-critical systems such as transportation electrification and medical equipment are particularly significant~\cite{ref1}. Another important characteristic of PMSMs is the requirement for accurate rotor position measurement to achieve efficient field-oriented control (FOC). Although sensorless control approaches are available, rotor position sensors remain the preferred solution in many applications. This preference primarily arises from the limitations of sensorless algorithms, particularly their reduced accuracy during transient operating conditions and the challenges associated with reliable startup performance~\cite{ref4},~\cite{ref5}. Among various control strategies, FOC, also referred to as vector control, provides superior dynamic performance and control accuracy~\cite{ref2}. However, its effectiveness depends heavily on accurate rotor position and current measurements, which ensure proper alignment and regulation of the stator magnetic field with respect to the rotor field. Therefore, accurate rotor magnetic field position information is essential for effective field orientation in PMSMs~\cite{ref6}. An overview of the FOC implementation block diagram for torque control applications is shown in Fig.~1. 

\begin{figure}[H]
    \centering
    \includegraphics[width=1.0\linewidth]{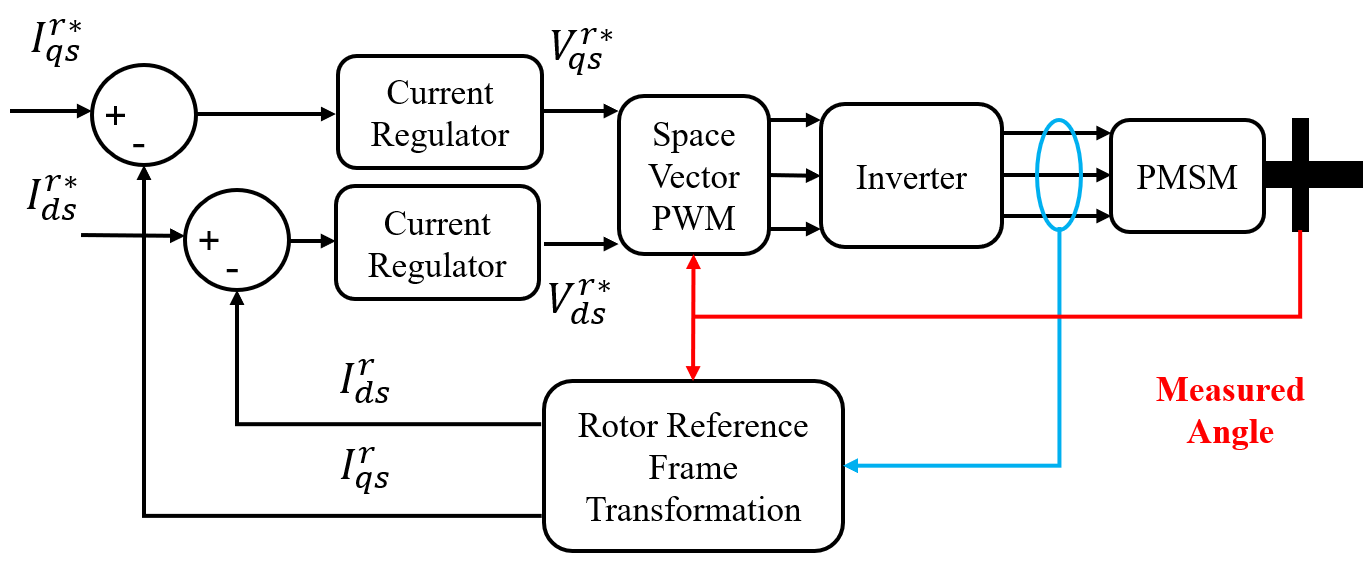}
    \caption{Block Diagram of a Field-Oriented Controller for a PMSM.}
    \label{fig:placeholder}
\end{figure}

In this implementation, the measured phase currents are transformed into the rotor reference frame to determine the required voltage commands for torque control. These voltages are subsequently transformed back into the stator reference frame to drive the motor. Both the forward and inverse coordinate transformations utilize the instantaneous rotor position to ensure correct alignment of the stator flux through voltage control.

Torque accuracy requirements, noise tolerance, space constraints, EMI tolerance, cost, and reliability requirements are factors that contribute to the selection of a position sensor for a particular application. Resolvers, magnetic Hall sensors, and encoders are some of the key technologies available for rotor position sensing in Sensed-FOC implementations for PMSMs. Regardless of the sensor type, the precision of the acquired signals is critical to ensure accurate field orientation and reliable torque regulation~\cite{ref6}. However, considering the mechanical interfacing required between the sensor and the motor, errors in measurement may be unavoidable~\cite{ref7}. Other means of error may be due to manufacturing flaws, EMI, poor design, or faulty sensors~\cite{ref4}. Thus, sensor calibration is essential, and it is typically performed for each motor drive combination in a given application prior to deployment.

Position sensor errors in PMSMs can be categorized into two types: Position Sensor Offset Error (PSOE) and Position Sensor Harmonic Error (PSHE)~\cite{ref4}. Equation~(1) represents both of the aforementioned errors in mathematical form, using Fourier decomposition. The measured rotor position $\theta_m$ can be expressed as the true rotor angle $\theta_r$ combined with a constant offset $\Delta\theta$ which accounts for static misalignment, and a series of sinusoidal distortions corresponding to different harmonic orders. Formally, the error terms can be described as in~(1)~\cite{ref4}.
\begin{equation}
\theta_m = \theta_r + \Delta\theta + \sum_{k=1}^{n} A_k \sin\!\left( k\theta_r + \phi_k \right)
\label{eq:position_error}
\end{equation}
Here, the integer $k$ denotes a specific harmonic order, $A_k$ is the amplitude of the $k^{\mathrm{th}}$ harmonic, and $\phi_k$ its corresponding phase angle. The PSOE ($\Delta\theta$) refers to the constant or the static misalignment between the sensor zero position and the actual rotor flux vector.

Depending on their severity, these errors may degrade torque accuracy, increase torque ripple, elevate acoustic noise, and compromise overall system integrity through induced mechanical vibrations~\cite{ref4}. Hence, mitigating such errors either through sensor calibration or real-time compensation is essential to ensure system efficiency and long-term reliability. The effects of PSOE and methods for compensating this failure mode and strategies for adaptive correction have been widely studied in~\cite{ref16,ref17,ref18}. However, active and adaptive compensation of PSHE has received limited attention, mainly due to the complex nature of harmonics. Thus, the next sub-sections review existing literature on harmonic compensation, followed by a novel approach for real-time adaptive harmonic compensation.

\subsection{Literature Review on Existing Harmonic Compensation Methods}

Numerous approaches have been studied and presented for position sensor harmonic error measurement and correction. These methods are primarily passive in nature, limiting the ability to actively regulate the harmonic error that may change under various operating conditions. Authors in~\cite{ref8} and~\cite{ref9} investigated periodic ripples in the analog rotor position caused by offset and gain mismatches between the sine and cosine outputs of sinusoidal encoders. Building on this, the present work analyzes these effects using a rotating coordinate system based on the $dq$ transformation, where the proposed compensator directly employs the synchronous $d$-axis component to detect the amplitudes of offset and gain errors. The proposed method is primarily limited to correcting nonlinear position sensor errors arising from offset and gain mismatches. Additionally, the method relies on accurate $dq$ transformation and synchronous reference frame analysis, making its performance sensitive to rotor position estimation errors and coordinate transformation inaccuracies, which may reduce the overall compensation effectiveness. The author in~\cite{ref10} proposed a novel adaptive compensation approach based on reference and adaptive models. In this method, the reference model generates the fundamental current vector, while the adaptive model estimates the harmonic current vector. The position error information is extracted through the cross product of the reference and adaptive current vectors. This method achieves real-time compensation of periodic position errors by continuously updating the output of the adaptive model to match the reference model. However, the method assumes that the nonlinear position error only consists of a single harmonic component, which may not accurately represent all practical harmonic error scenarios. The method presented in~\cite{ref11} utilizes a Kalman filter to estimate the true joint position by combining a dynamic model of the direct-drive actuator, an applied actuator command, and measurements from the uncalibrated sensor. Based on the estimated true position, a lookup table is generated to compensate for sensor errors and improve measurement accuracy. However, this method is model-dependent, as it relies on an accurate system dynamics model and Kalman filter tuning, which may not be valid under parameter variations or unmodeled dynamics. Additionally, the use of a look-up table (LUT) makes the compensation largely dependent on previously learned conditions, which may limit its adaptability to changing operating environments or nonlinear error variations over time. In~\cite{ref12}, a lookup table-based interpolation method is introduced, where encoder signal issues such as offsets, mismatched amplitudes, and waveform distortion are directly corrected. The dependency of this method on the LUT results in limited adaptability. The author in~\cite{ref13} proposed an observer that utilizes the resolver’s sinusoidal signals along with measured data to estimate motor speed and position, while minimizing the angle error to accurately determine the rotor’s current position. In practice, angle tracking observers (ATOs) often lose accuracy at very low speeds~\cite{ref14}. Furthermore, the authors in~\cite{ref15} proposed an adaptive algorithm to compensate for harmonics in the position signal of PMSMs. However, this method primarily focuses on correcting harmonics with single components only.

In summary, existing methods demonstrate drawbacks such as model dependency, limited ability to correct harmonics, and/or more importantly inability to effectively adapt to changing harmonic content in real-time.  As a result, existing approaches still face significant challenges in achieving fully effective compensation of position signal harmonics in PMSM drives in a model-free, active, and adaptive manner. The proposed PSHE compensation method addresses this requirement by eliminating dependence on motor parameters and implementing a self-regulating control mechanism based on an extremum seeking algorithm that effectively suppresses harmonic components in the position signal, thereby enhancing the robustness, efficiency, and precision of PMSM operation. The most commonly used method for correcting harmonics in the position signal is the lookup table approach, and its limitations are discussed in the following section.

\subsection{Position Sensor Harmonics Correction using Lookup Table Method}
In the look-up table (LUT)-based method, position measurements prior to error correction are collected from the existing sensor, and the corresponding error signals with respect to the actual rotor position are identified. These error characteristics are then used to construct the lookup table for harmonic error compensation in measured position (Fig.~2). The error between the measured position and the actual position can be expressed as the harmonic components present in the measured position signal as,
\begin{equation}
E(\theta_r) = \sum_{k=1}^{n} A_k \sin\left(k\theta_r + \phi_k\right)
\label{eq:2}
\end{equation}
Despite the error being defined as a function of true rotor position, in reality the input to the LUT is the measured position (which contains error). There are two key forms of error that emerge in this form of implementation as shown in Fig.~2. First is the non-linearity error when position is injective and a more severe form of error when the position signal is non-injective, as discussed below. The error term observed by the LUT can be represented as follows.
\begin{equation}
E(\theta_m) = \sum_{k=1}^{n} A_k \sin\left(k\theta_m + \phi_k\right)
\label{eq:3}
\end{equation}
Hence, the corrected estimate of the rotor position using the LUT-based method can be expressed as:
\begin{equation}
\theta_c = \theta_m - E(\theta_m)
\label{eq:4}
\end{equation}
The following discussion mathematically establishes the condition required for non-injectivity to occur in the LUT-based position harmonic correction approach, which leads to a more severe form of distortion in the position signal. Consider the derivative of the measured position signal with respect to the true rotor position, as given in (1).
\begin{equation}
\frac{d\theta_m}{d\theta_r}
=
1+\sum_{k=1}^{n} kA_k \cos(k\theta_r+\emptyset_k)
\label{eq:29}
\end{equation}
For accurate error mapping, the corresponding function must be one-to-one, which requires the derivative of the measured position signal to remain strictly positive to ensure monotonicity. Thus, the following condition should be satisfied.
\begin{equation}
1+\sum_{k=1}^{n} kA_k \cos\left(k\theta_r+\emptyset_k\right) > 0
\label{eq:30}
\end{equation}
Therefore, the following need to be satisfied for non-injectivity.
\begin{equation}
\sum_{k=1}^{n} kA_k \cos\left(k\theta_r + \phi_k\right) \leq -1
\label{eq:30}
\end{equation}
The above result demonstrates that, given a harmonic error, the non-injectivity is governed by the combination of $k$, $A_k$, $\theta_r$, and $\phi_k$. Depending on the harmonic content present in the position signal, the condition governing monotonic behavior may vary. However, when the condition~(7) is violated, the mapping loses monotonicity, leading to the formation of turning points. This results in multiple values of $\theta_m$ corresponding to the same error value, thereby giving rise to the non-injectivity issue in the LUT-based implementation. Fig.~2 illustrates two distinct sources of error associated with the LUT-based method. In Case~1, the error arises during error compensation, since the actual rotor position $\theta_r$ is unavailable during operation, and therefore the LUT is indexed using $\theta_m$ instead of $\theta_r$. Consequently, for a given incoming measured position value at point A, the LUT returns the error term corresponding to point B, whereas the correct error term should correspond to point C, which represents the true rotor position at that time instant. This mismatch during LUT indexing becomes the dominant source of residual compensation error as derived analytically. In Case~2, the non-injectivity issue manifests during error correction. When the harmonic amplitude becomes sufficiently large, the measured position signal becomes non-monotonic, causing the same incoming $\theta_m$ value to map to the same LUT location, even though it should be mapped to different locations, as illustrated in the figure.
\begin{figure}[H]
    \centering
    \includegraphics[width=1.0\linewidth]{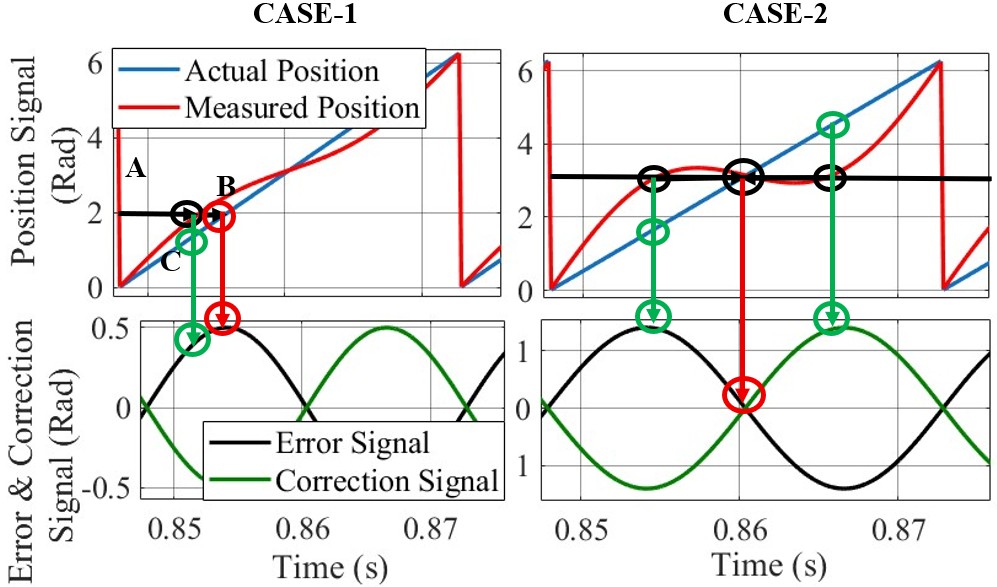}
    \caption{Errors Associated with the LUT-Based Harmonic Error Correction Method for PMSMs.}
    \label{fig:placeholder}
\end{figure}
Apart from the drawbacks discussed above, the LUT-based method is non-adaptive to variations in the harmonic characteristics of the position signal. This limitation arises because the LUT is typically constructed offline under fixed harmonic conditions; therefore, any variation in the amplitude, phase, or spectral content of the position error cannot be accommodated without re-identifying or reconstructing the entire lookup table. Consequently, the method lacks robustness under varying operating conditions and requires repeated recalibration to maintain compensation accuracy. The experimental validation of the aforementioned errors is presented in Section~IV.

\section{\textsc{Extremum Seeking Control for PSHC Compensation}}

Extremum Seeking Control (ESC) is an adaptive, model-free optimization algorithm that locates the extremum of a unimodal nonlinear map by introducing a perturbation, extracting gradient information from the system response, and iteratively updating the control input to ensure convergence to the optimal operating condition~\cite{ref20}. The application of ESC requires a unimodal objective function with a well-defined extremum, which ensures convergence of the system input to the corresponding optimal operating point. Leveraging this property, the following section describes the use of ESC to minimize harmonic errors in the position signal by adaptively driving the system toward a zero-harmonic condition using measurable signals, without requiring explicit knowledge of the harmonic components. Furthermore, it discusses the selection of the appropriate objective function, its unimodality, stability, convergence, and how its extremum leads to the elimination of harmonic components.

\subsection{Theoretical Framework for formulating the objective function and Unimodal Characteristics}

This section formulates the objective function for compensating harmonic components in the position signal using the Extremum Seeking Control (ESC) method. Considering a salient PMSM, the electromagnetic torque equation may be expressed as in~(8),
\begin{equation}
T_{em} = \left(\frac{3}{2}\right)\left(\frac{P}{2}\right)\left[\lambda_{m}^{\prime r} I_{qs}^{r} + \left(L_d - L_q\right) I_{qs}^{r} I_{ds}^{r}\right]
\label{eq:6}
\end{equation}
For a non-salient machine where $L_d \approx L_q$, or a salient machine with $I_{ds}^{r} = 0\,$ leads to the electromagnetic torque relation as follows,
\begin{equation}
T_{em} = \left(\frac{3}{2}\right)\left(\frac{P}{2}\right)\lambda_{m}^{\prime r} I_{qs}^{r}
\label{eq:7}
\end{equation}
According to~\cite{ref6}, the relationship between PSOE and electromagnetic torque, expressed in terms of the commanded q-axis current reference, can be represented as in~(10), assuming $I_{ds}^{r} = 0$.
\begin{equation}
T_{em} = \left(\frac{3}{2}\right)\left(\frac{P}{2}\right)\lambda_{m}^{\prime r} I_{qs}^{r\ast} \cos(\Delta \theta)
\label{eq:8}
\end{equation}
Here $\Delta\theta$, which represents the PSOE, can be defined as the combined effect of all harmonic components together with a static PSOE in the position signal.
\begin{equation}
\Delta \theta = \Delta \theta_{\text{static}} + \sum_{k=1}^{n} A_k \sin\left(k\theta_r + \phi_k\right)
\label{eq:9}
\end{equation}
The effect of static PSOE using ESC has been previously investigated in~\cite{ref1}. Since the primary focus of this work is harmonic error compensation, the static PSOE is assumed to be zero by considering perfect alignment between the sensor zero position and the rotor flux vector.
\begin{equation}
\Delta \theta = \sum_{k=1}^{n} A_k \sin\left(k\theta_r + \phi_k\right)
\label{eq:10}
\end{equation}
Then the electromagnetic torque can be rewritten as follows:
\begin{equation}
T_{em} = \frac{3P}{4}\lambda_{m}^{\prime r} I_{qs}^{r\ast}
\cos\!\left(\sum_{k=1}^{n} A_k \sin(k\theta_r + \phi_k)\right)
\label{eq:11}
\end{equation}
The above analysis indicates that the harmonic components must be zero to eliminate the torque ripple associated with the position harmonics. Furthermore, it can be shown that the q-axis voltage applied to a field-oriented controlled PMSM with a clockwise-rotating PSOE is given by~(14)~\cite{ref17}.
\begin{equation}
V_{qs}^{r} = r_s I_{qs}^{r\ast} + \lambda_{m}^{\prime r} \omega_r \cos(\Delta \theta)
\label{eq:12}
\end{equation}
As explained above, $\Delta\theta = \sum_{k=1}^{n} A_k \sin(k\theta_r + \phi_k)$, and $V_{qs}^{r}$ can be rewritten as follows:
\begin{equation}
V_{qs}^{r} = r_s I_{qs}^{r\ast} + \lambda_{m}^{\prime r} \omega_r
\cos\!\left(\sum_{k=1}^{n} A_k \sin\!\left(k\theta_r + \phi_k\right)\right)
\label{eq:13}
\end{equation}
According to~\cite{ref4}, the harmonic signals are considered as small-signal components ($A_k < 1$). The higher-order terms can be neglected based on the Taylor series expansion of $\cos x$:
\begin{equation*}
\cos\!\left(\sum_{k=1}^{n} A_k \sin(k\theta_r + \phi_k)\right)
\approx
1 - \sum_{k=1}^{n} \frac{A_k^2}{2}\sin^2(k\theta_r + \phi_k)
\end{equation*}
\begin{equation}
V_{qs}^{r}
= r_s I_{qs}^{r\ast}
+ \lambda_{m}^{\prime r}\omega_r
\left(
1 - \sum_{k=1}^{n}\frac{A_k^2}{2}\sin^2(k\theta_r + \phi_k)
\right)
\label{eq:15}
\end{equation}
\begin{equation}
V_{qs}^{r}
= r_s I_{qs}^{r\ast}
+ \lambda_{m}^{\prime r}\omega_r
- \lambda_{m}^{\prime r}\omega_r
\sum_{k=1}^{n}\frac{A_k^2}{2}\sin^2(k\theta_r + \phi_k)
\label{eq:16}
\end{equation}
Let's consider $h = \sum_{k=1}^{n} A_k \sin(k\theta_r + \phi_k)$; hence $h^2$ can be derived as follows:
\begin{multline}
h^2 = \left[\sum_{k=1}^{n} A_k \sin\left(k\theta_r + \phi_k\right)\right]^2
= \sum_{k=1}^{n} A_k^2 \left[\sin\left(k\theta_r + \phi_k\right)\right]^2 \\
+ 2 \sum_{j \neq k}^{n} A_k A_j
\sin\left(j\theta_r + \phi_j\right)\sin\left(k\theta_r + \phi_k\right)
\label{eq:17}
\end{multline}
Given that the harmonic amplitudes are $A_k < 1$, the product $A_k A_j$ is significantly smaller than unity, and thus the cross terms can be neglected. Hence, (18) can be simplified as follows:
\begin{equation*}
h^2 = \left[\sum_{k=1}^{n} A_k \sin\!\left(k\theta_r + \phi_k\right)\right]^2
\approx
\sum_{k=1}^{n} A_k^2 \sin^2\!\left(k\theta_r + \phi_k\right)
\end{equation*}
Accordingly, (17) can be reformulated as follows:
\begin{equation}
V_{qs}^{r} = r_s I_{qs}^{r\ast} + \lambda_{m}^{\prime r} \omega_r - \frac{\lambda_{m}^{\prime r} \omega_r}{2} h^2
\label{eq:19}
\end{equation}
From~(19), it is evident that $V_{qs}^{r}$ is a unimodal function of the harmonic content in the position signal, denoted by $h$. The function attains its extremum (maximum) value of $r_s I_{qs}^{r\ast} + \lambda_{m}^{\prime r} \omega_r$ when $h = 0$. Hence, it is clear that using the above defined objective function attains its extremum when harmonics in the system converge to zero.

\subsection{Extraction of the Harmonic Content in the Position Signal for Compensation}

This section presents the mathematical formulation of the harmonic content in the position signal. As indicated by the unimodal function in (19), the presence of harmonic content in the system is necessary for analysis. However, the harmonic components in the position signal are not directly accessible. The following derivation establishes the approach on how the signals that exist within FOC are being used to capture harmonic content, subsequently to be fed to the ESC for perfect cancellation. According to [4], \(V_{ds}^{r}\) for a field-oriented control-based current-regulated non-salient PMSM can be written as:
\begin{equation}
V_{ds}^{r}
=
-\omega_r L_q I_{d_{\mathrm{ref}}}
-
\sin(\Delta\theta)\,\omega_r \lambda_{m}^{\prime r}
\label{eq:20}
\end{equation}
Using (12), equation (20) can be rewritten as follows:
\begin{equation}
V_{ds}^{r}
=
-\omega_r L_q I_{d_{\mathrm{ref}}}
-
\sin\!\left(
\sum_{k=1}^{n}
A_k \sin(k\theta_r + \phi_k)
\right)
\omega_r \lambda_{m}^{\prime r}
\label{eq:21}
\end{equation}
Assuming the harmonic signals are considered as small-signal components for analysis simplification, (21) can be reformulated as follows:
\begin{equation}
\sin\!\left(
\sum_{k=1}^{n} A_k \sin(k\theta_r + \phi_k)
\right)
\approx
\sum_{k=1}^{n} A_k \sin(k\theta_r + \phi_k)
\label{eq:22}
\end{equation}
\begin{equation}
V_{ds}^{r}
=
-\omega_r L_q I_{d_{\mathrm{ref}}}
-
\left(
\sum_{k=1}^{n} A_k \sin(k\theta_r + \phi_k)
\right)
\omega_r \lambda_{m}^{\prime r}
\label{eq:23}
\end{equation}
From (23), it can be observed that \(V_{ds}^{r}\) is influenced by the harmonics present in the measured position signal, sharing the same frequencies, and its magnitude is directly proportional to the amplitude of each corresponding harmonic. Hence, (23) can be rewritten as follows:
\begin{equation}
V_{ds}^{r}
=
-\omega_r L_q I_{d_{\mathrm{ref}}}
-
\sum_{k=1}^{n} A_k^{\prime} \sin(k\theta_r + \phi_k)
\label{eq:24}
\end{equation}
Fig.~3 presents simulation results of \(V_{ds}^{r}\) and the injected harmonic components, which validate the mathematical proof.
\begin{figure}[H]
    \centering
    \includegraphics[width=0.95\linewidth]{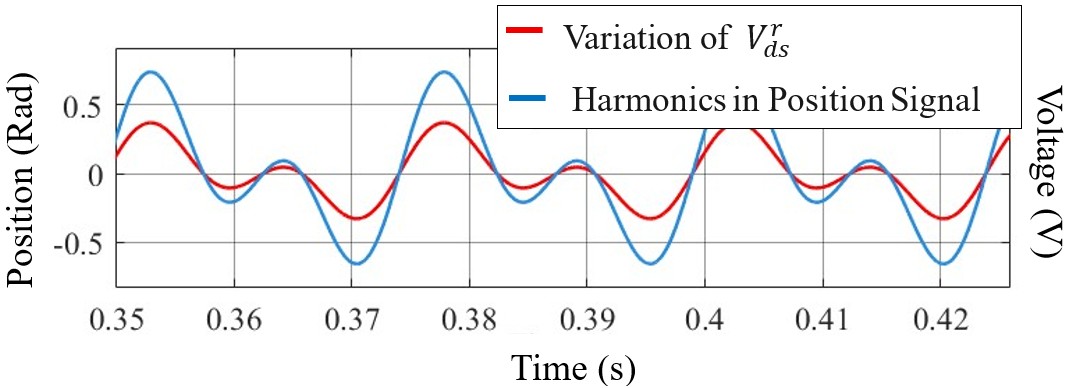}
    \caption{Simulation results of the response of \(V_{ds}^{r}\) and the injected harmonic.}
    \label{fig:3}
\end{figure}
\subsection{Mathematical Formulation for Harmonic Correction}
The following discussion presents the theoretical foundation for correcting position measurements distorted by harmonic components in a field-oriented controlled PMSM drive system. Given that a signal proportional to the harmonics is available in (24), the scaling of the correction term is implemented through the ESC, which suppresses the majority of the position error adaptively. As shown in (25), the corrected position signal \(\theta_c\) incorporates both the measured rotor position and the harmonic correction terms, as expressed below:
\begin{equation}
\theta_c
=
\theta_m
-
A \sum_{k=1}^{n} A_k^{\prime} \sin(k\theta_r + \phi_k)
\label{eq:25}
\end{equation}
Here, ``A'' denotes the perturbation amplitude associated with the harmonic content in the position signal. The harmonic component of the position signal, given by \(\sum_{k=1}^{n} A_k^{\prime}\sin(k\theta_r + \phi_k)\), is obtained after removing the DC component from \(V_{ds}^{r}\). Consequently, using (1), equation (25) can be rewritten as:
\begin{equation}
\begin{aligned}
\theta_c
&= \theta_r + \sum_{k=1}^{n} A_k \sin(k\theta_r + \phi_k) \\
&\quad - A \sum_{k=1}^{n} A_k^{\prime} \sin(k\theta_r + \phi_k)
\end{aligned}
\label{eq:26}
\end{equation}
\begin{equation}
\theta_c
=
\theta_r
+
\left(A_k - A A_k^{\prime}\right)
\sum_{k=1}^{n} A_k \sin(k\theta_r + \phi_k)
\label{eq:27}
\end{equation}
As explained above, the ESC algorithm reaches its maximum value when the harmonic content in the system converges to zero. Accordingly, \(A_k - A A_k^{\prime}\) converges to zero, and the corrected position \(\theta_c\) converges to the true rotor position \(\theta_r\). Fig.~4 presents the open-loop simulation results illustrating the unimodal behavior of \(V_{qs}^{r}\) under amplitude perturbation \((A)\), thereby supporting the above conclusion on unimodal behavior of the function. Subplot 1 shows the variation of \(V_{qs}^{r}\) during the correction process with injected multiple harmonics. Subplot 2 illustrates the corresponding variation in the position signal throughout the correction process. Subplot 3 depicts the error, defined as \((A_k - A A_k^{\prime})\), representing the difference between the actual harmonics and the applied correction signal.
\begin{figure}[H]
    \centering
    \includegraphics[width=0.85\linewidth]{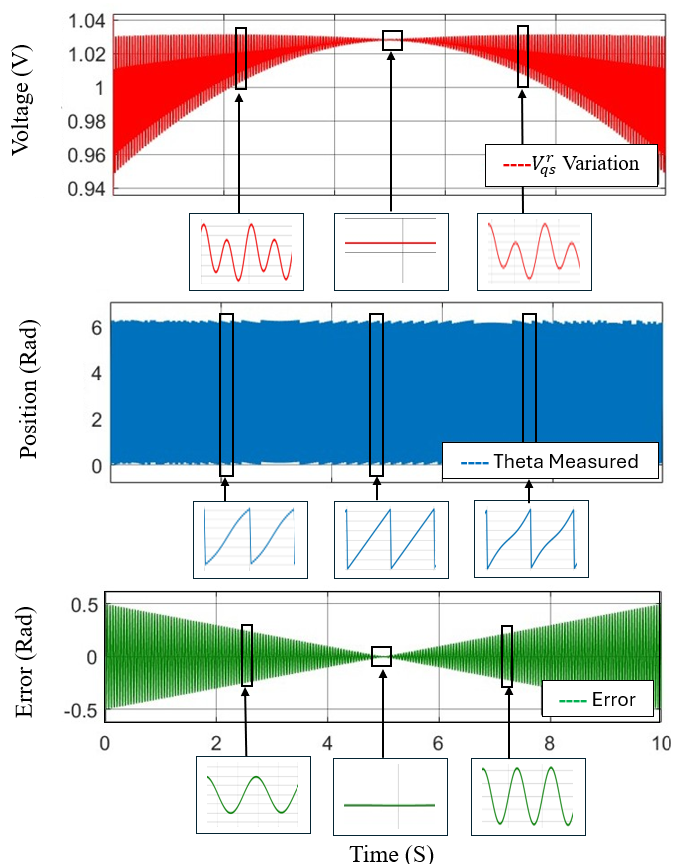}
    \caption{Variation of \(V_{qs}^{r}\), Measured Position, and Error during the Correction Process Under Amplitude Perturbation.}
    \label{fig:4}
\end{figure}
As observed, the maximum value of \(V_{qs}^{r}\) occurs when the correction signal amplitude matches the existing harmonic component in the position signal, corresponding to \(A_k = A A_k^{\prime}\), which is consistent with the theoretical derivation presented above.

\subsection{Stability, Convergence, and Parameter Selection Criteria in ESC Algorithm}

The stability of the sinusoidal ESC scheme employed in this work has been rigorously established in~\cite{ref20}. Using averaging theory and singular perturbation analysis, the ESC system can be represented by an average dynamic model in which the convergence behavior is governed by the Jacobian of the averaged system, as illustrated in (28). According to the averaging-based stability analysis, stable convergence toward the extremum operating point is achieved when the eigenvalues of the averaged Jacobian remain negative through appropriate selection of the ESC parameters. A generalized block diagram of the ESC is illustrated in Fig.~5.
\begin{figure}[H]
    \centering
    \includegraphics[width=0.83\linewidth]{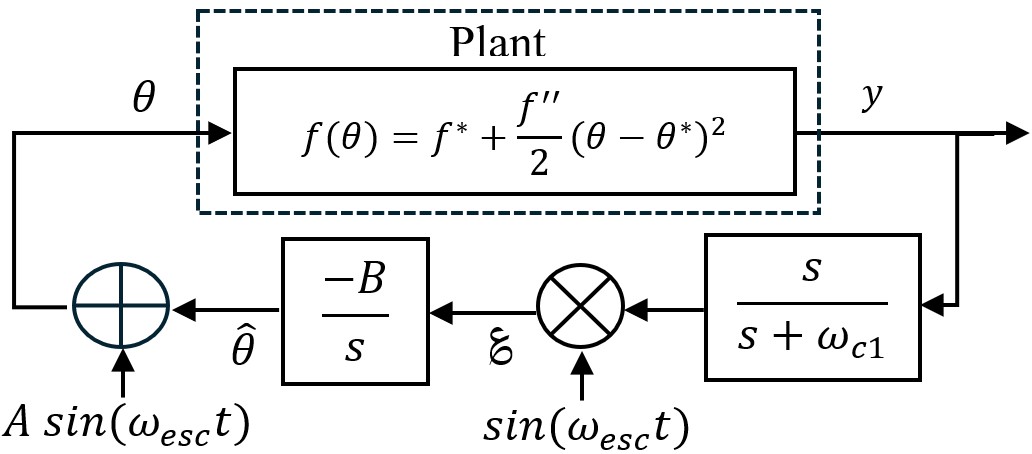}
    \caption{Generalized Block Diagram of the ESC~\cite{ref20}.}
    \label{fig:5}
\end{figure}
\begin{equation}
J_{\mathrm{avg}} =
\begin{bmatrix}
-\frac{B A f^{\prime\prime}}{2\omega_{\mathrm{esc}}} & 0 \\
0 & -\frac{\omega_{c1}}{\omega_{\mathrm{esc}}}
\end{bmatrix}
\label{eq:28}
\end{equation}
From the average Jacobian, it can be observed that the term \(-\omega_{c1}/\omega_{\mathrm{esc}}\) is always negative for positive values of \(\omega_{c1}\) and \(\omega_{\mathrm{esc}}\). However, the sign of the remaining eigenvalue depends on the term \(f^{\prime\prime}\), which represents the curvature of the objective function around the extremum point. Therefore, the stability of the ESC system can be controlled through the appropriate selection of the integrator gain ``B''. By properly choosing the sign of ``B'' according to the sign of \(f^{\prime\prime}\), all poles of the averaged system can be placed in the left-half plane, thereby ensuring stable convergence of the ESC algorithm.

The proposed extremum-seeking-based harmonic error mitigation algorithm is integrated into the field-oriented control framework, making parameter selection a critical task. In addition to the current regulator bandwidth, four parameters must be tuned: \(\omega_{\mathrm{esc}}\), \(\omega_{c1}\), \(A\), and \(B\). The current regulator bandwidth \(\omega_{bw}\) is determined by the application requirements and is independent of the ES loop. However, its value must still be considered when selecting certain ES parameters. Among them, the perturbation frequency \(\omega_{esc}\) governs both the speed of the search process and the rate of convergence. In practice, \(\omega_{esc}\) is typically chosen within the range of \(0.1\,\omega_{bw}\) to \(0.5\,\omega_{bw}\)~\cite{ref1}. Selecting a smaller \(\omega_{esc}\) increases the convergence time and necessitates the use of a high-pass filter with a lower cutoff frequency. Conversely, the upper bound on \(\omega_{esc}\) is imposed to prevent attenuation of the perturbation signal as it propagates through the current regulator. The cutoff frequency \(\omega_{c1}\) is chosen to keep most of the AC search signal while removing the DC part of the signal. Parameter ``A'' represents the amplitude of the small perturbation. As noted in~\cite{ref19}, this perturbation is amplified by the rotor speed, such that fluctuations in ``A'' cause variations in motor current and, consequently, increased torque ripple. The integrator gain ``B'' influences the ESC convergence time. A large ``B'' value may cause undesirable overshoots during the search process, whereas a very small value slows convergence. In this application, practical results were obtained with ``B'' in the range of 1 to 50, allowing it to be tuned according to the convergence requirements of a specific application~\cite{ref1}.

\subsection{Proposed ESC Approach}

In the proposed ESC approach, the initial perturbation is injected into the corrected position signal in the form of the harmonic being compensated. The amplitude of \(V_{ds}^{r}\) (after removing the DC component), which is proportional to the actual harmonics present in the position signal, is perturbed to compensate for the existing harmonics in the position signal. The resulting harmonically perturbed position signal is then propagated through the FOC algorithm, which indirectly introduces a corresponding perturbation in \(V_{qs}^{r}\). The perturbed \(V_{qs}^{r}\) entering the ESC high-pass filter contains both the slope information of \(V_{qs}^{r}\) and a DC offset. The high-pass filter in the ESC loop extracts only the AC component, which carries the slope information of \(V_{qs}^{r}\) with respect to the PSHE. In the subsequent demodulation stage of the ESC algorithm, the AC component is extracted and processed to obtain the slope information, which is then fed into the integrator. The integrator accumulates this value until the extremum point is reached. At the extremum condition, the slope becomes zero; therefore, the ESC algorithm naturally converges to the optimal operating point that effectively cancels the position sensor harmonic error and remains there until a change in the harmonic error occurs. Fig.~6 illustrates the block diagram of the Self-Regulating Position sensor harmonic compensation using the ESC algorithm for PMSM Drives.

\begin{figure*}
 \centering
  \includegraphics[width=0.93\textwidth]{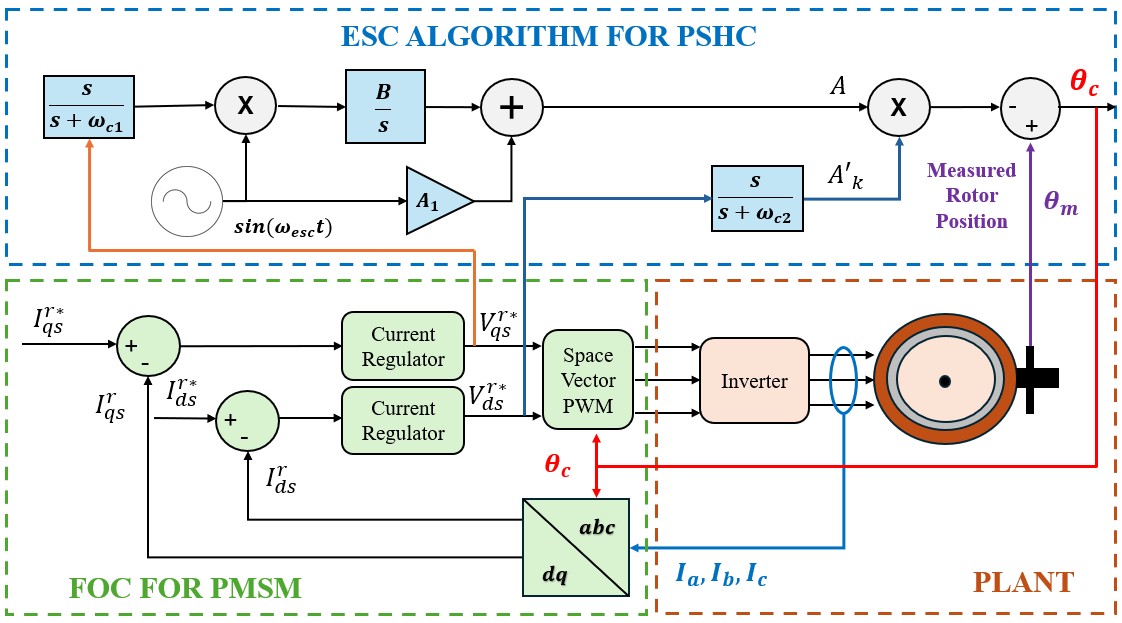}
\caption{Block Diagram of the Self-Regulating Position Sensor Harmonic Compensation using ESC Algorithm for PMSM Drive.}
\label{fig:esc_block}
\end{figure*}

\section{Experimental Results}
The proposed harmonic compensation method was experimentally validated on a dSPACE MicroLabBox platform interfaced with a PMSM–DC motor dynamometer setup. The parameter values of the PMSM and the ESC-based PSHC are listed in Table~I.
\begin{table}[H]
\caption{\textsc{PMSM and ESC Parameters for the Experimental Setup}}
\label{tab:table1}
\centering
\begin{tabular}{|c|c|}
\hline
Parameter & Value \\
\hline
Phase Resistance $r_s$ & $71.4~\mathrm{m\Omega}$ \\
\hline
q-axis Inductance $L_q$ & $226.8~\mu\mathrm{H}$ \\
\hline
Flux Linkage Constant $\lambda_m$ & $76~\mathrm{mVs/rad}$ \\
\hline
Controller Bandwidth $(\omega_{bw})$ & $20~\mathrm{rad/s}$ \\
\hline
ESC Integral Gain $(B)$ & $25$ \\
\hline
Perturbation Amplitude $(A_1)$ & $0.05$ \\
\hline
ESC Cutoff Frequency $(\omega_{C1})$ & $0.4~\mathrm{rad/s}$ \\
\hline
Cutoff Frequency $(\omega_{C2})$ & $0.4~\mathrm{rad/s}$ \\
\hline
ESC Search Frequency $(\omega_{esc})$ & $3~\mathrm{rad/s}$ \\
\hline
\end{tabular}
\end{table}
The experimental setup, as shown in Fig.~7, consists of a DC motor used for speed control, while the PMSM generates torque under field-oriented control (FOC) using encoder-based position feedback. A TIMKEN M15-1000-8-1-9-2-1-2 encoder with 1000 pulses per revolution (PPR) and 8 poles was employed for position sensing. Torque measurements were obtained using a FUTEK-TRS605 torque sensor.
\begin{figure}[H]
    \centering
    \includegraphics[width=0.90\linewidth]{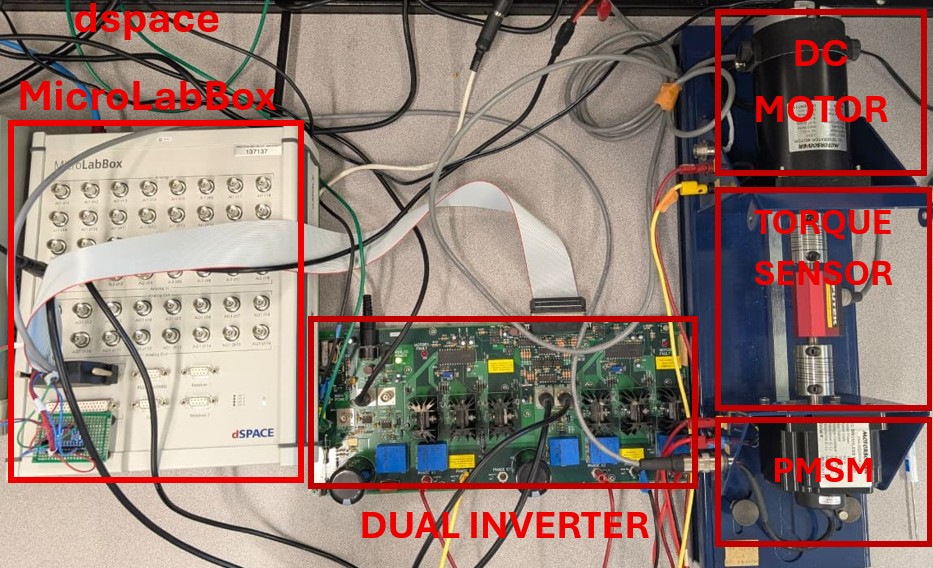}
    \caption{Experimental Setup Used for Validation of the PSHE Based on ESC}
    \label{fig:placeholder}
\end{figure}
To validate the proposed ESC-based harmonic correction algorithm under different harmonic conditions, the following harmonic signals are considered.
\begin{enumerate}
    \item $0.6\sin\left(\theta_r(t)\right) + 0.4\sin\left(2\theta_r(t)\right)$
    
    \item $0.6\sin\left(\theta_r(t)+0.17\right) + 0.4\sin\left(2\theta_r(t)\right)$
    
    \item $0.6\sin\left(\theta_r(t)+0.17\right) + 0.4\sin\left(2\theta_r(t)\right)+0.2\sin\left(3\theta_r(t)\right)$
\end{enumerate}

Figs.~8, 10, and 12 illustrate the variations in $I_{ds}^{r}$, $I_{qs}^{r}$, torque, and position response, divided into four operating regions. Region~1 (R1) represents the system operation without harmonic injection. Region~2 (R2) corresponds to the system response after the introduction of harmonic error while the ESC algorithm remains disabled. Region~3 (R3) represents the correction interval during which the ESC algorithm is activated, whereas Region~4 (R4) corresponds to the steady-state operation after compensation. All experimental test cases were conducted at a motor speed of $200~\mathrm{rpm}$ with $I_{ds}^{r}=0~\mathrm{A}$ and $I_{qs}^{r}=2~\mathrm{A}$. Figs.~9, 11, and 13 illustrate the time-domain and frequency-domain torque ripple profiles for Regions~R1, R2, and R4 of the test cases presented above.

\begin{figure}[H]
    \centering
    \includegraphics[width=0.9\linewidth]{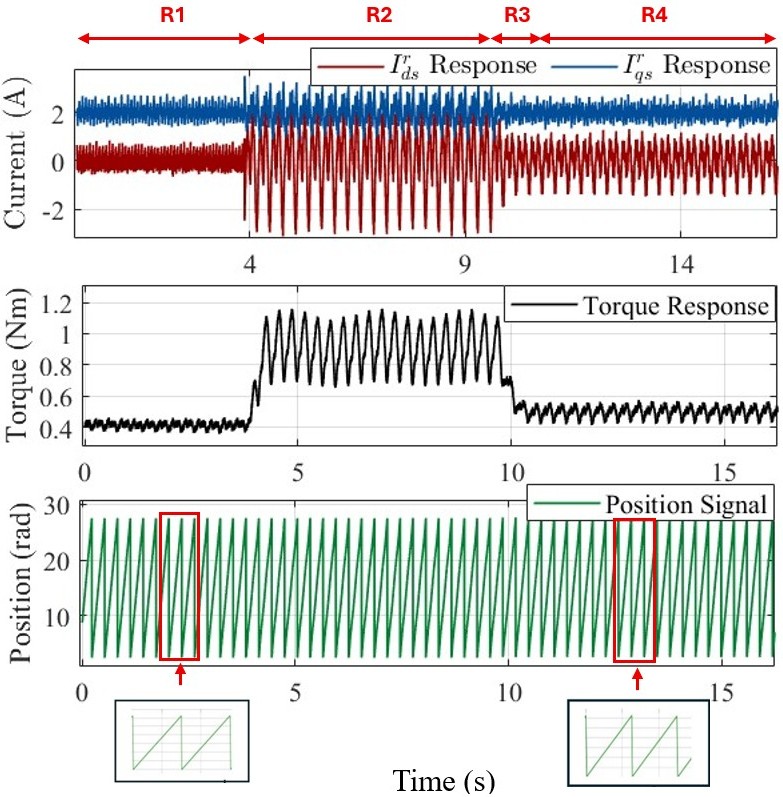}
    \caption{\(I_{ds}^{r}\), \(I_{qs}^{r}\), torque, and position signal variations across four regions for the harmonic case 1 $(I_{ds}^r = 0~\mathrm{A},\ I_{qs}^r = 2~\mathrm{A},\ \text{and speed} = 200~\mathrm{rpm})$}
    \label{fig:placeholder}
\end{figure}

\begin{figure}[H]
    \centering
    \includegraphics[width=0.95\linewidth]{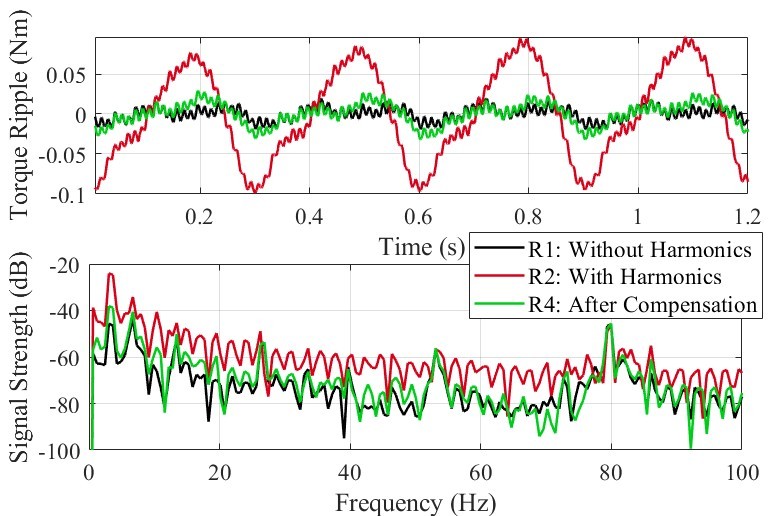}
    \caption{Torque spectrum in Regions 1, 2, and 4 for the harmonic case represented in Fig.~8.}
    \label{fig:placeholder}
\end{figure}

\begin{figure}[H]
    \centering
    \includegraphics[width=0.95\linewidth]{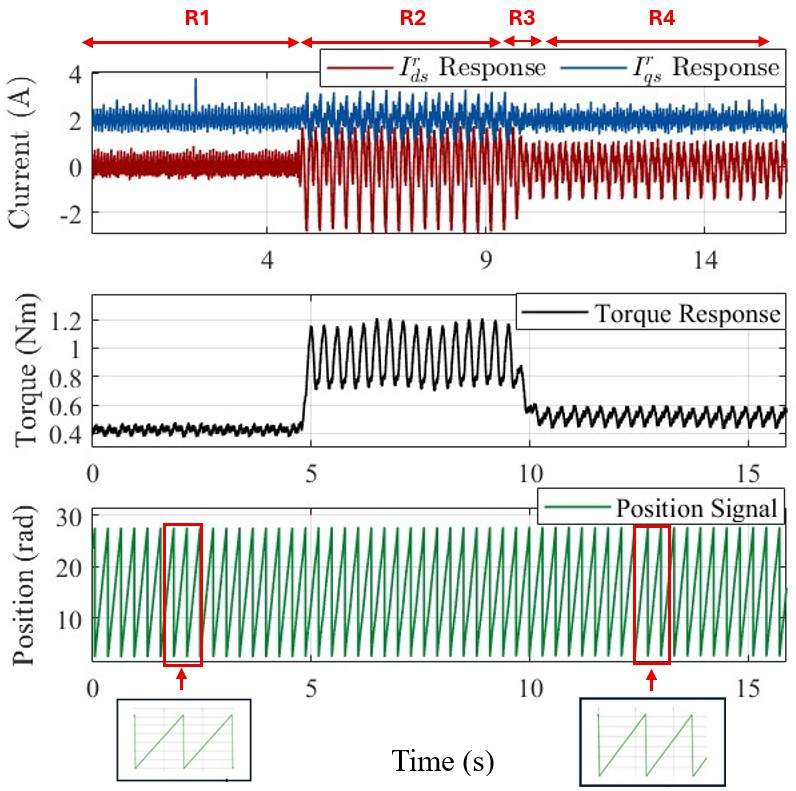}
    \caption{\(I_{ds}^{r}\), \(I_{qs}^{r}\), torque, and position signal variations across four regions for the harmonic case 2 $(I_{ds}^r = 0~\mathrm{A},\ I_{qs}^r = 2~\mathrm{A},\ \text{and speed} = 200~\mathrm{rpm})$}
    \label{fig:placeholder}
\end{figure}

\begin{figure}[H]
    \centering
    \includegraphics[width=0.95\linewidth]{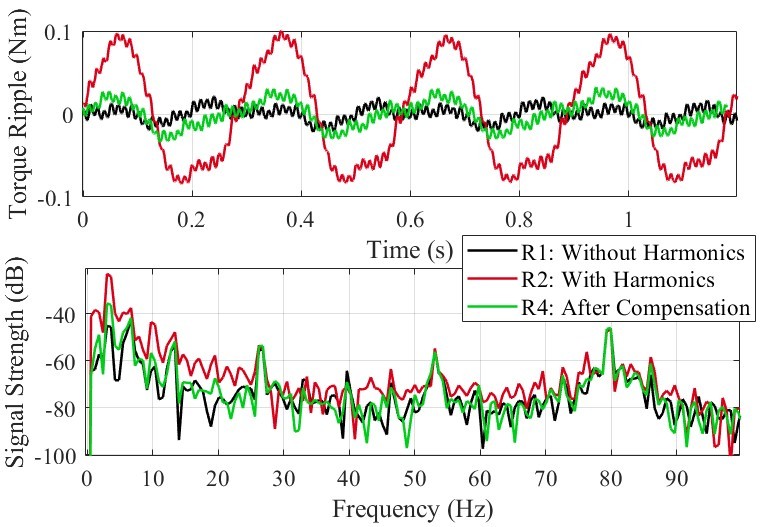}
    \caption{Torque spectrum in Regions 1, 2, and 4 for the harmonic case represented in Fig.~10.}
    \label{fig:placeholder}
\end{figure}

\begin{figure}[H]
    \centering
    \includegraphics[width=0.95\linewidth]{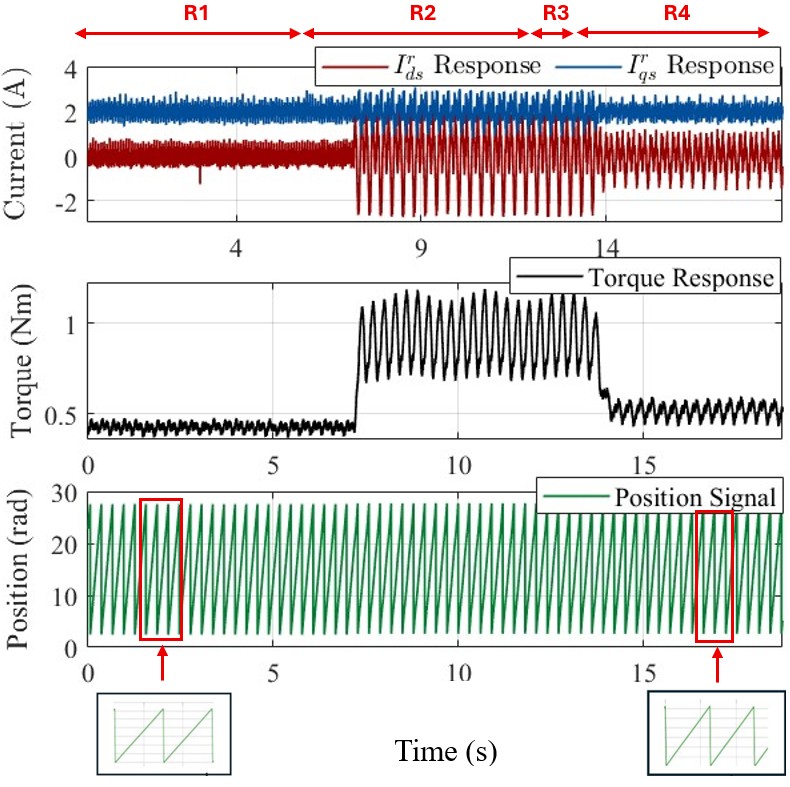}
    \caption{\(I_{ds}^{r}\), \(I_{qs}^{r}\), torque, and position signal variations across four regions for the harmonic case 3 $(I_{ds}^r = 0~\mathrm{A},\ I_{qs}^r = 2~\mathrm{A},\ \text{and speed} = 200~\mathrm{rpm})$}
\end{figure}

\begin{figure}[H]
    \centering
    \includegraphics[width=0.95\linewidth]{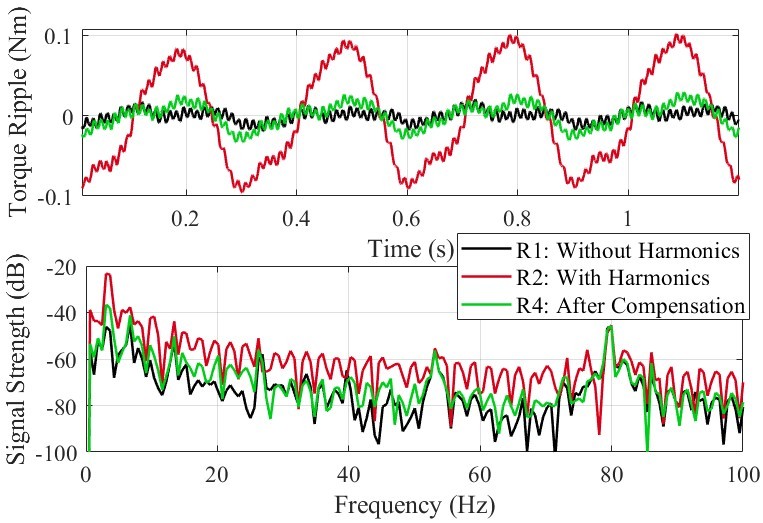}
    \caption{Torque spectrum in Regions 1, 2, and 4 for the harmonic case represented in Fig.~12.}
    \label{fig:placeholder}
\end{figure}

The figure above clearly illustrates the impact of harmonics in the position signal on the torque response, as well as the effectiveness of the proposed system in attenuating the torque ripple induced by harmonics in the position signal.

Multiple experiments were conducted under various operating conditions to assess the effectiveness of the proposed method across different speeds and current levels. Table~II outlines the test cases conducted to evaluate the effectiveness of the position harmonic correction using ESC.

\begin{table}[H]
\caption{\textsc{Test Cases Evaluated Under Different Operating Conditions Using the ESC-Based Harmonic Compensation Method}}
\label{tab:table2}
\centering
\begin{tabular}{|c|c|c|c|}
\hline
Harmonic Signal & Speed 100 rpm & Speed 300 rpm & Speed 400 rpm \\
\hline
$0.6\sin(\theta_r(t))$ 
& $I_{ds}^{r} = 0~\mathrm{A}$ 
& $I_{ds}^{r} = 0~\mathrm{A}$ 
& $I_{ds}^{r} = 0~\mathrm{A}$ \\
$+~0.4\sin(2\theta_r(t))$
& $I_{qs}^{r} = 0.5~\mathrm{A}$ 
& $I_{qs}^{r} = 1.5~\mathrm{A}$ 
& $I_{qs}^{r} = 2~\mathrm{A}$ \\
\hline
\end{tabular}
\end{table}

Fig.~14 through Fig.~16 illustrate the variations in \(I_{ds}^{r}\), \(I_{qs}^{r}\), torque, and the corresponding position response during the algorithm’s correction period for the test cases described above. Fig.~14 shows the responses of $I_{ds}^r$, $I_{qs}^r$, torque, and position signal variations across four regions for $I_{ds}^r = 0~\mathrm{A}$, $I_{qs}^r = 0.5~\mathrm{A}$, and a motor speed of $100~\mathrm{rpm}$.

\begin{figure}[H]
    \centering
    \includegraphics[width=0.92\linewidth]{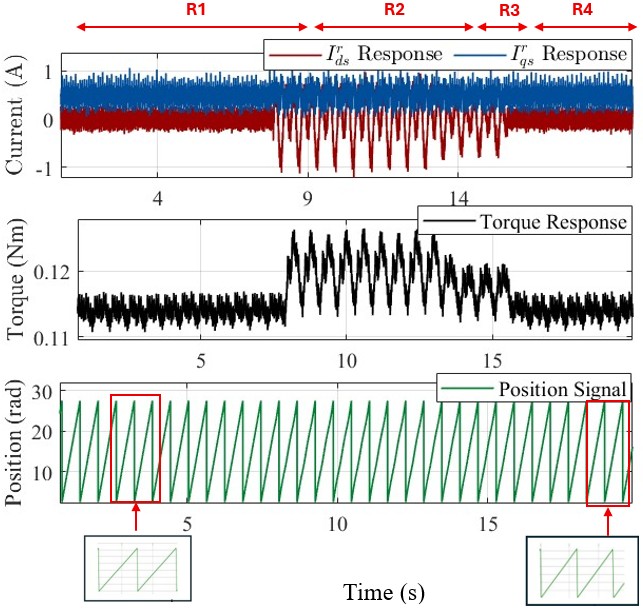}
    \caption{Response of \(I_{ds}^{r}\), \(I_{qs}^{r}\), torque, and position for the first test case in TABLE II $(I_{ds}^r = 0~\mathrm{A},\ I_{qs}^r = 0.5~\mathrm{A},\ \text{and speed} = 100~\mathrm{rpm})$}
    \label{fig:placeholder}
\end{figure}
Fig.~15 shows the responses of $I_{ds}^r$, $I_{qs}^r$, torque, and position signal variations across four regions for $I_{ds}^r = 0~\mathrm{A}$, $I_{qs}^r = 1.5~\mathrm{A}$, and a motor speed of $300~\mathrm{rpm}$.
\begin{figure}[H]
    \centering
    \includegraphics[width=0.92\linewidth]{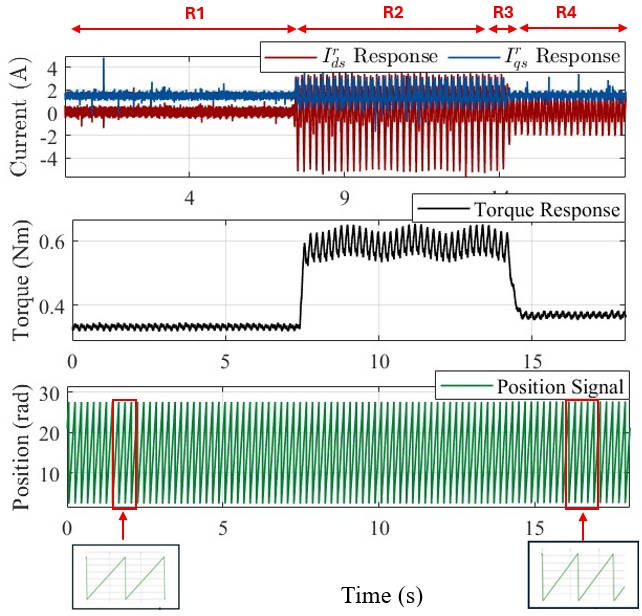}
    \caption{Response of \(I_{ds}^{r}\), \(I_{qs}^{r}\), torque, and position for the second test case in TABLE II $(I_{ds}^r = 0~\mathrm{A},\ I_{qs}^r = 1.5~\mathrm{A},\ \text{and speed} = 300~\mathrm{rpm})$}
    \label{fig:placeholder}
\end{figure}

Fig.~16 shows the responses of $I_{ds}^r$, $I_{qs}^r$, torque, and position signal variations across four regions for $I_{ds}^r = 0~\mathrm{A}$, $I_{qs}^r = 2~\mathrm{A}$, and a motor speed of $400~\mathrm{rpm}$.

\begin{figure}[H]
    \centering
    \includegraphics[width=0.92\linewidth]{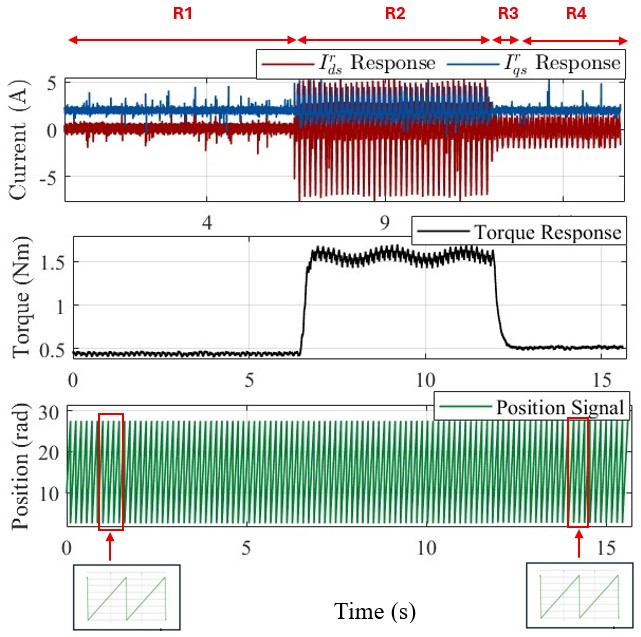}
    \caption{Response of \(I_{ds}^{r}\), \(I_{qs}^{r}\), torque, and position for the third test case in TABLE II $(I_{ds}^r = 0~\mathrm{A},\ I_{qs}^r = 2~\mathrm{A},\ \text{and speed} = 400~\mathrm{rpm})$}
    \label{fig:placeholder}
\end{figure}

The following figures illustrate the impact of harmonic injection on the system response. When harmonics are introduced into the position signal, both $I_{ds}^r$ and $I_{qs}^r$ exhibit noticeable oscillations, while the torque response contains significant ripple components along with a DC offset. The presence of the DC offset arises from the static offset indirectly introduced by position harmonics as shown in equation (13). As the proposed correction algorithm converges to the actual harmonic amplitude, the oscillatory components in $I_{ds}^r$ and $I_{qs}^r$ are substantially suppressed. Consequently, the torque ripples, including the associated DC offset, are significantly reduced.These results validate the effectiveness of the proposed algorithm under varying speed, torque, and harmonic conditions.

\section{Harmonic Compensation Using The Traditional Look-up Table Method}

Under this section, harmonic compensation is implemented using the lookup table (LUT) method to demonstrate the advantage of the ESC-based method proposed in this paper. The compensation performance is evaluated under three different harmonic scenarios. For each test case, a dedicated LUT is employed to compensate for the harmonic components in the position signal. The following harmonic signals are considered to demonstrate the operation of the proposed method.
\begin{enumerate}
    \item $0.3\sin\left(\theta_r(t)\right) + 0.2\sin\left(2\theta_r(t)\right)$
    
    \item $0.6\sin\left(\theta_r(t)\right) + 0.4\sin\left(2\theta_r(t)\right)$
    
    \item $0.9\sin\left(\theta_r(t)\right) + 0.6\sin\left(2\theta_r(t)\right)$
\end{enumerate}

Fig.~17 presents the following subplots: Subplot~1 shows the position error between the corrected signal and the actual position; Subplot~2 shows the corrected position signal; Subplot~3 shows the position error spectrum ; Subplot~4 shows the torque ripple; and Subplot~5 shows the torque ripple spectrum for each test case.

\begin{figure}[H]
    \centering
    \includegraphics[width=0.95\linewidth]{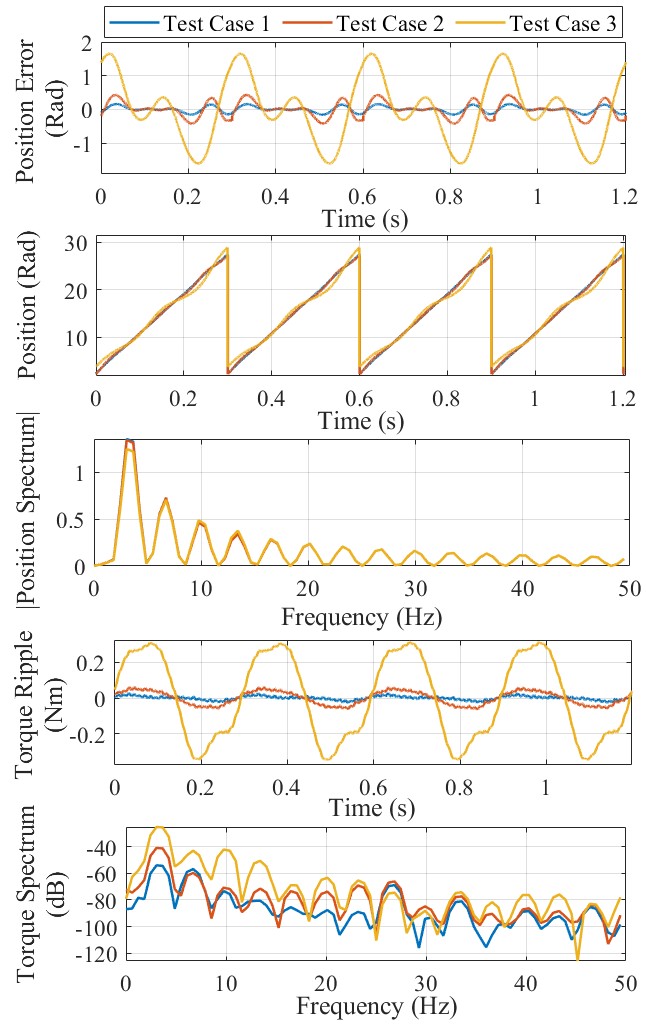}
    \caption{Response of the position correction and torque characteristics of the PMSM under the evaluated test cases using the LUT method.}
    \label{fig:placeholder}
\end{figure}

Under LUT-based harmonic compensation, the corrected position exhibits residual nonlinear errors even when an exact lookup table is employed for correction. These errors arise primarily from inaccurate interpolation within the lookup table, as discussed in Section~I-B.

\section{Accuracy Evaluation of the Proposed Approach and LUT-Based Method}

In this section, the accuracy of the corrected position signal is evaluated based on the theoretical derivations and validated using the experimental results presented in Figs.~8, 10, and 12. Furthermore, the accuracy evaluation of the LUT-based method is also analyzed using the experimental results presented in Fig.~17, enabling a comparative assessment between the two approaches. As described in (27), when the ESC algorithm converges to its extremum, the term $(A_k - A A_k^{\prime})$ approaches zero, and the corrected position $\theta_c$ theoretically converges to the true rotor position $\theta_r$. However, in practice, a slight deviation from this ideal convergence is observed due to non-idealities in the experimental setup. The error distribution, along with the corresponding maximum error with respect to the corrected position for the three experimental test cases presented in Figs.~8, 10, and 12, is illustrated in Fig.~18.
\begin{figure}[H]
    \centering
    \includegraphics[width=0.96\linewidth]{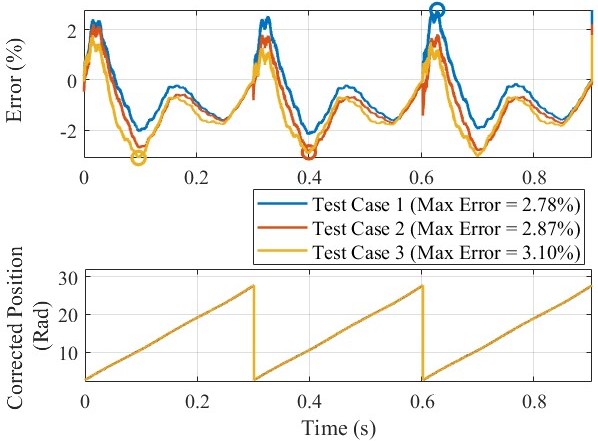}
    \caption{Error distribution and the corresponding maximum error with respect to the corrected position for the experimental test cases presented in Figs.~8, 10, and 12.}
    \label{fig:placeholder}
\end{figure}
Fig.~19 illustrates the error distribution and the corresponding maximum error with respect to the corrected position for the test cases presented in Fig.~17.
\begin{figure}[H]
    \centering
    \includegraphics[width=0.96\linewidth]{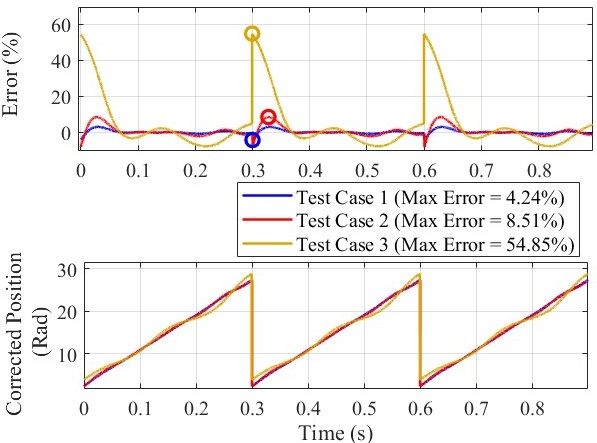}
    \caption{Error distribution and the corresponding maximum error with respect to the corrected position for the experimental test cases presented in Fig.~17.}
    \label{fig:placeholder}
\end{figure}
When comparing the accuracy of both methods, it is evident that the proposed ESC-based adaptive compensation method achieves a lower error percentage and demonstrates improved adaptability compared to the conventional LUT-based method. Nevertheless, the proposed ESC-based method is developed based on the small-signal approximation condition $(A_k < 1)$~\cite{ref4}, which represents a limitation of the proposed approach. In situations where this condition is violated, the ESC algorithm may not converge exactly to the optimum point required for complete harmonic compensation. However, the experimental results for Case~3, as illustrated in Fig.~12, demonstrate that the proposed method is capable of achieving nearly $97\%$ error correction despite the presence of comparatively large combined harmonic amplitudes. These results indicate the robustness and practical effectiveness of the proposed method under severe multi-harmonic operating conditions.

\section{Conclusion}

Position sensor harmonic errors in field-oriented controlled Permanent Magnet Synchronous Machines (PMSMs) are known to induce torque ripple, leading to mechanical vibrations that can result in premature component failure. Conventional compensation strategies, such as look-up table-based methods, are often limited by their inability to adapt to varying harmonic levels. To address this, this paper proposes an adaptive motor position harmonic compensation strategy based on Extremum Seeking Control (ESC). The work provides a comprehensive analytical framework for the ESC-based approach, followed by experimental validation across various harmonic profiles and operating conditions. The results demonstrate the system's ability to effectively suppress position sensor harmonics in real-time. Additionally, the accuracy of the proposed method is discussed to provide a balanced assessment for practical implementation.

\section*{Acknowledgments}
The authors gratefully acknowledge the support of the U.S. National Science Foundation for this research, provided through the NSF EAGER Grant (No. 2440506).

\section*{Biography Section}

\begin{IEEEbiography}[{\includegraphics[width=1.1in,height=1.25in,clip,keepaspectratio]{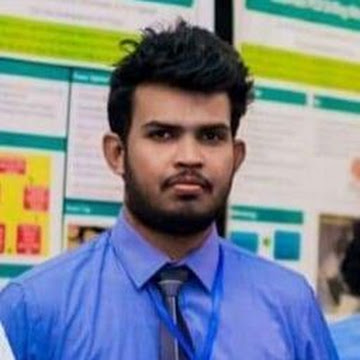}}]{Gayan Dissanayake} 
(Student Member, IEEE) received the B.Sc. degree in Electrical and Electronic Engineering from the Faculty of Engineering, University of Peradeniya, Sri Lanka, in 2020. Following graduation, he served as an instructor in the same department. He is currently pursuing a Ph.D. degree in Electrical Engineering at Western Michigan University, USA, where he is with the Transportation Electrification and Applied Mechatronics Laboratory. His research interests include electrical machines and drives, power electronics, power systems, microgrids, and renewable energy systems.
\end{IEEEbiography}

\begin{IEEEbiography}[{\includegraphics[width=1in,height=1.25in,clip,keepaspectratio]{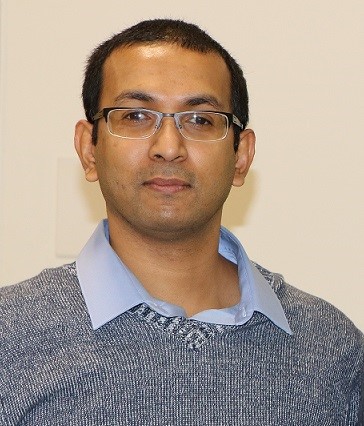}}]{Sandun Kuruppu} 
(M’14–SM’19) received his B.S. degree in Electrical and Electronics Engineering from University of Peradeniya, Sri Lanka in 2007 and M.S. and Ph.D. degrees from Purdue University, West Lafayette, Indiana, USA, in 2010 and 2013, respectively.

He is an Associate Professor of Electrical and Computer Engineering at Western Michigan University (WMU), Kalamazoo, Michigan, USA, and the director of the Transportation Electrification and Applied Mechatronics Lab. Prior to his current position, he was with Saginaw Valley State University, Nexteer Automotive, Texas Instruments Kilby Labs, and Delphi Electronics and Safety. His current research interests include fault prognostics, diagnostics, localization, mitigation in mechatronic systems, power electronics, vehicle stability control, and extremum seeking controls for traction applications.

\end{IEEEbiography}

\end{document}